\documentclass[aps,prb,amssymb,twocolumn,showpacs,supersriptaddress,groupeaddress]{revtex4-1}
\usepackage{amssymb}
\usepackage[percent]{overpic}
\usepackage{graphicx}
\usepackage{mwe}
\usepackage{braket}
\usepackage{mathtools, amssymb, amsmath}
\usepackage{cancel}
\usepackage{float}
\usepackage{mathrsfs}
\usepackage{gensymb}

\usepackage{amsmath}
\usepackage{graphicx}
\usepackage{dcolumn}
\usepackage{bm}
\usepackage{subcaption}
\usepackage{bm}
\usepackage{natbib,hyperref}
\usepackage{hyperref}
\usepackage{color}
\hypersetup{colorlinks=true, 
    linkcolor=red,          
    citecolor=magenta,        
    filecolor=gree,      
    urlcolor=blue           
}

\usepackage{float}

\begin{document}

\title{Landau-Zener-St\"uckelberg Interferometry in dissipative Circuit Quantum Electrodynamics}

 \author{Mariano Bonifacio$^{1}$, Daniel Dom\'inguez$^{1}$ and Mar\'ia Jos\'e S\'anchez$^{1,2}$}

\affiliation{$^{1}$Centro At\'omico Bariloche and Instituto Balseiro, \\ 
	R8402AGP San Carlos de Bariloche, Argentina.}
\affiliation{$^{2}$Instituto de Nanociencia y Nanotecnolog\'{\i}a (INN),CONICET-CNEA, Argentina.}

\begin{abstract}

We study Landau-Zener-St\"uckelberg (LZS) interferometry in a cQED architecture under effects of dissipation. To be specific, 
we consider  a superconducting qubit driven by a dc+ac signal and coupled to a transmission line resonator, but our results are valid for general qubit-resonators devices. To take the environment into account, we assume that the resonator is coupled to an ohmic quantum bath. The Floquet-Born-Markov master equation is numerically solved to obtain the dynamics of the system for arbitrary amplitude of the drive and different time scales. We unveil important differences in the resonant patterns between the Strong Coupling  and Ultra Strong Coupling regimes in the qubit-resonator interaction, which are mainly due to the magnitude of photonic gaps in the energy spectrum of the system. We identify in the LZS patterns the contribution of the qubit gap and the photonic gaps, showing that for large driving amplitudes the patterns present a weaving structure due to the combined intercrossing of the different gaps contributions.

\end{abstract}


\maketitle
 \section{Introduction}
 
Circuit Quantum Electrodynamics (cQED) \cite{blais_2004,wallraff_2004,xiang_2013} -the study of the interaction between superconducting circuits behaving as artificial atoms and transmission line resonators- has become  one of the test beds for quantum information processing tasks \cite{gu_2017}. 
Originally implemented for studying on-chip light-matter interactions, the enormous advances during the last decade in the development of long-lived qubits-resonators devices, have shown the possibility  of performing a large number of high-fidelity quantum gates, entangling  and coupling distant qubits to realize two qubit gates and to carry out non-demolition readout operations \cite{esteve_2012,dicarlo_2009, dicarlo_2010,campagne-ibarcq_2018, paik_2011, johnson_2010, walter_2017, van_loo_2013, eichler_2012, didier_2015}.
 
Landau-Zener-St\"uckelberg (LZS) interferometry  has been established as a powerful tool  to probe the energy level spectrum of a superconducting qubit  and to study coherent phenomena  for large driving amplitudes \cite{shevchenko_2010}. In typical LZS protocols, the
qubit energy levels are modulated back and forth through
an avoided crossing at a frequency  faster than the qubit decoherence
rate.  Strong driving dynamic
has been experimentally investigated in superconducting qubits \cite{oliver_2005,berns_2006,berns_2008,oliver_2009,izmalkov_2008} and quantum dots devices \cite{sillanpaa_2006,wilson_2007,petta_2010,ribeiro_2013,forster_2014}.
In addition, LZS interferometry was recently proposed  to determine relevant information related to the coupling of a qubit with a noisy environment  
\cite{forster_2014,blattmann_2015,mi_2018,ferron_2012,ferron_2016,gramajo_2019}.

In the present work, we analize LZS conditions in cQED, by  strongly  driving a  qubit
coupled to  a quantum mode of an oscillator.  We  take into account the coupling of the system to a quantum bath and  study  the dissipative dynamics in the  strong  qubit-resonator coupling using the  Floquet-Markov master equation \cite{grifoni_1998,kohler_1998, ferron_2016}. We focus in the strong driving regime (large driving amplitudes),  beyond  standard  approaches that restrict the driven dynamics to the rotating wave approximation (RWA). In this way  we analyze the emergence of the  multi-Floquet modes in the dissipative scenario of cQED, unvealing  the interference patterns  and population features that are not captured within the RWA.
Although we  use   parameters typical for superconducting qubits and  mircrowave resonators experiments \cite{gustavsson_2012}, our study can be  extended to
analyze  Floquet spectrocopic experiments recently implemented in  driven qubits coupled to  mechanical resonators  and for high or low-frequency driving fields
\cite{kervinen_2019}.

The paper is organized as follows: In section \ref{sec:I} we introduce the driven cQED model  Hamiltonian for the case of a flux qubit driven by an ac flux. We   discuss the structure of the energy spectrum in the absence of driving, which will be useful  to understand
the emergence of multi-Floquet modes under the driving protocol.
Section \ref{sec:II} is devoted to analyze the LZS interference patterns neglecting dissipation, with the aim of  comparing   the patterns that emerge  due  to  the photonic gaps in the  driven Jaynes Cummings model valid under the RWA, with those of the full driven Rabi Hamiltonian, where the counter rotating terms in the qubit-resonantor interaction are taken into account. 
In  Sec. \ref{sec:III} we extend the analysis to the dissipative case, which is relevant for realistic experimental situations. The strong coupling (SC) and ultrastrong coupling (USC) in the qubit-resonator interaction are analyzed in detail and the structure of the respective LZS patterns are characterized for finite times and  in the stationary regime, after full relaxation with the bath degrees of freedom. A concluding summary is provided  in Sec. \ref{sec:IV}.

\section{Model Hamiltonian}
\label{sec:I}
 We consider a flux qubit driven by an ac harmonic flux  and  coupled capacitively to  a transmission line resonator that contains one mode of the EM field,
as customary  in cQED architectures \cite{gustavsson_2012,yang_2016}.
The corresponding model Hamiltonian is 
\begin{equation}
H(t)=H_q{(t)}+H_r+H_{qr}\;, 
\label{eq:Hamil}
\end{equation}
where
\begin{equation}
        \begin{split}
            &H_q{(t)}=\frac{1}{2}\left[\varepsilon(t)\sigma_z+\Delta\sigma_x\right],\\ &H_r=\omega_r a^\dagger a,\\
            &H_{qr}= g \sigma_y \left(a+a^\dagger\right),\\
       \end{split}
    \label{eq:H_Rabi}
\end{equation}
are the   terms for the  driven flux qubit restricted to a two-level system \cite{ferron_2016}, the resonator and the Rabi interaction Hamiltonians, respectively. We take $\hbar=1$ in this work.
The qubit is driven by a time dependent bias $\varepsilon(t)= \varepsilon_0+A\cos{\omega t}$,
where $\varepsilon_0$ is the static bias component  on top of which is the harmonic ac modulation of amplitude $A$ and frequency $\omega$ \cite{izmalkov_2008, shevchenko_2010, ferron_2016, gramajo_2019}.  
The operators $\sigma_x$, $\sigma_y$ and $\sigma_z$ are the Pauli matrices and $\Delta$ is the coupling strength between the states $\ket{\uparrow}$ and $\ket{\downarrow}$ of the computational basis of the qubit.
In the absence of driving,  the qubit Hamiltonian $H_q$ has eigenenergies
$\pm\omega_q/2$, with  $\omega_q=\sqrt{{\varepsilon_0}^2+\Delta^2}$.
The frequency of the resonator is $\omega_r$  and  $a^\dagger$ ($a$) is the creation (annihilation) operator for resonator photons.   
 The capacitive coupling between the flux qubit  and the resonator here studied \cite{gustavsson_2012,yang_2016} is  represented in terms of the $\sigma_y$ operator, being  $g$  the coupling strength. 
 Other  well studied cases, like   a charge qubit coupled capacitively to a resonator \cite{blais_2004,wallraff_2004} or a flux qubit coupled inductively to a  resonator \cite{chiorescu2_2004} are modeled in terms of the operator  $\sigma_z$ in $H_{qr}$.
For the composite Hilbert space $\mathcal{H}_{qubit}\otimes \mathcal{H}_{resonator}$ we use the product state basis $\{\ket{\downarrow,n}, \ket{\uparrow,n}\}_{n\in \mathbb{N}}$, where $n$ is the eigenvalue of the  resonator  photon number
operator $a^\dagger a$.

Throughout this work we will consider  $\Delta/\omega_r=0.0038$,
which  corresponds to typical experiments in driven flux qubits with small qubit gap \cite{oliver_2005}, where $\Delta\sim 10-50$MHz,
while typical cavity frequencies are in the range of $\omega_r/2\pi\sim 10$GHz \cite{blais_2004}. In spite of these specific parameters, our results can be easily extended to other types of superconducting qubits. We study different  values of the coupling parameter in the range $g/\omega_r=0.0019 - 0.12$. 
In cQED it is customary to define  the Strong Coupling (SC) and Ultra Strong Coupling (USC) regimes, with the conditions $g\lesssim0.1\omega_{q/r}$ and $g>0.1\omega_{q/r}$, respectively \cite{niem_2010,gu_2017}. The USC regime has been experimentally achieved in recent years with superconducting qubits enabling the study of exciting and novel phenomena in the field of light-matter interaction \cite{gu_2017,yan_2018,forn_2019}.

\begin{figure}[h]
	\begin{subfigure}[ht!]{0.5\textwidth} 
		\centering
		\begin{overpic}[width=0.85\textwidth]{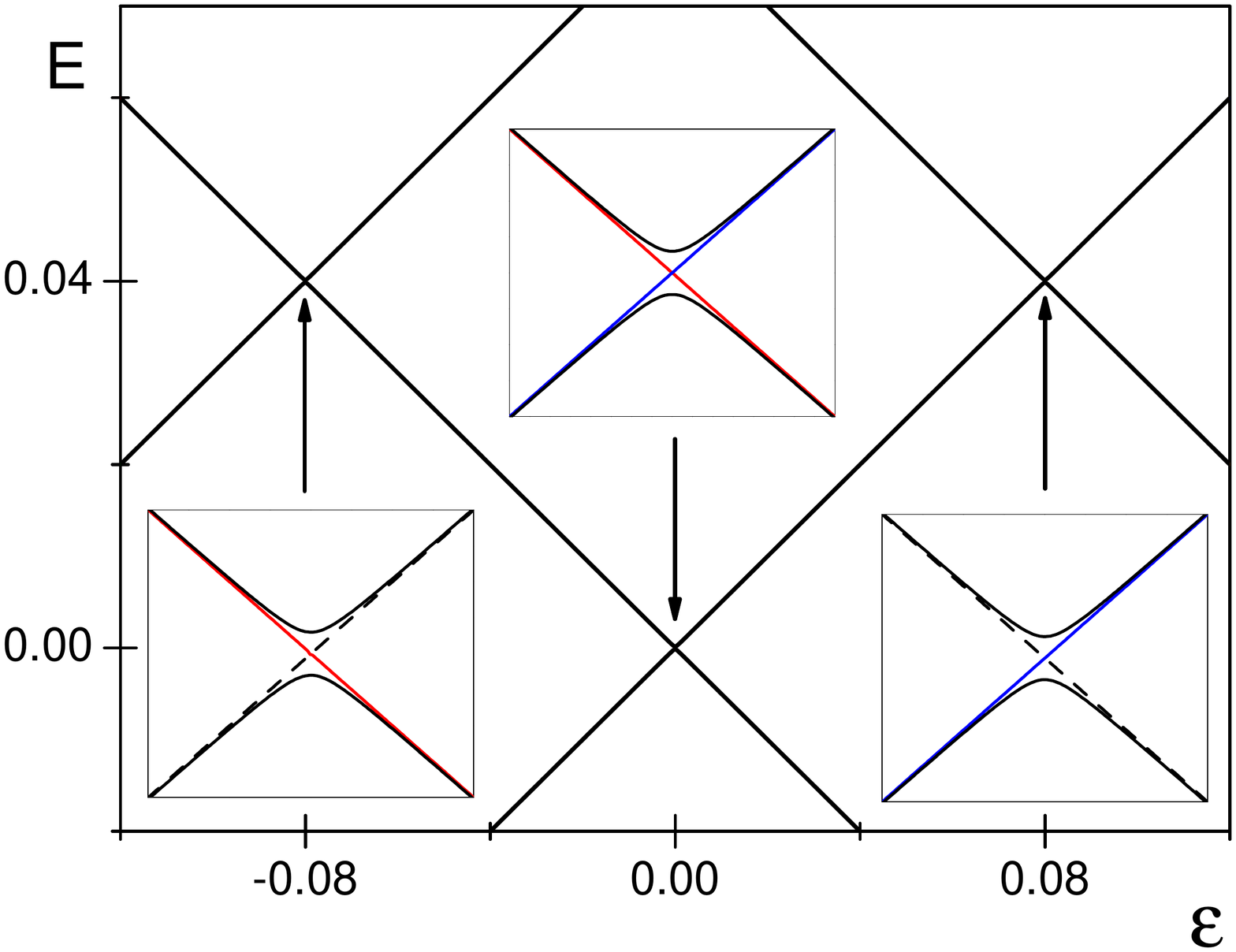}
			\put(3,58){(a)}
			\put (33,23.5) {\tiny$\displaystyle\ket{\uparrow,1}$}
			\put (20,23.5) {\tiny$\displaystyle\ket{\downarrow,0}$}
			\put (54.5,47) {\tiny$\displaystyle\ket{\uparrow,0}$}
			\put (42.5,47) {\tiny$\displaystyle\ket{\downarrow,0}$}
			\put (78,23.5) {\tiny$\displaystyle\ket{\uparrow,0}$}
			\put (65,23.5) {\tiny$\displaystyle\ket{\downarrow,1}$}
		\end{overpic}
		\label{fig:1a}
	\end{subfigure}
	\begin{subfigure}[ht!]{0.5\textwidth} 
		\centering
		\begin{overpic}[width=0.85\textwidth]{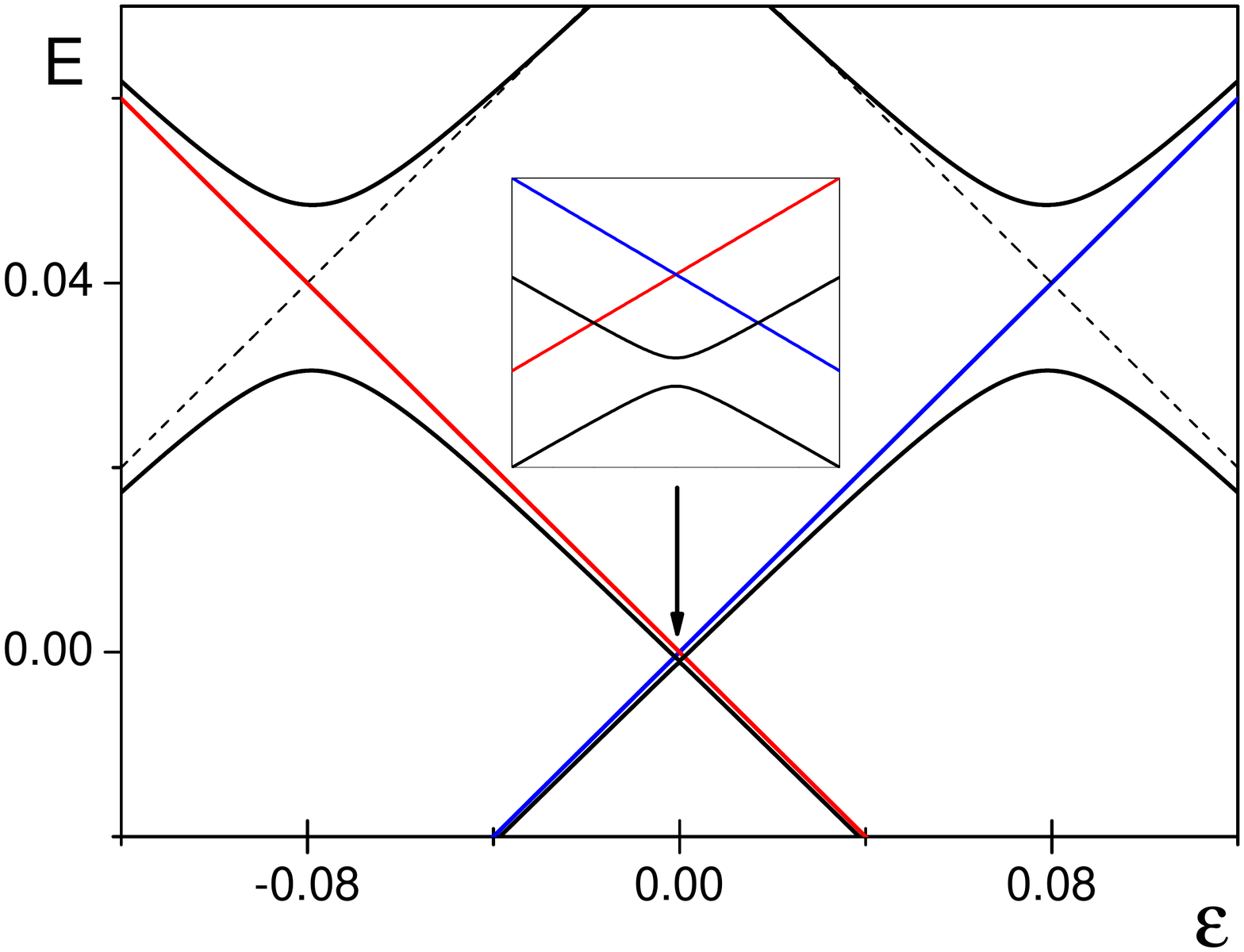}
			\put(3,58){(b)}
			\put (58,18) {\tiny$\displaystyle\ket{\downarrow,0}$}
			\put (39,18) {\tiny$\displaystyle\ket{\uparrow,0}$}
			\put (19,42.5) {\tiny$\displaystyle\ket{\uparrow,1}$}
			\put (79,42.5) {\tiny$\displaystyle\ket{\downarrow,1}$}
		\end{overpic}
		\label{fig:1b}
	\end{subfigure}
	\caption{Lowest energy levels of the Rabi Hamiltonian Eq.(\ref{eq:Hamil}) without driving ($A=0$), as a function of the dc bias $\varepsilon=\varepsilon_0$, for the parameters $\Delta/\omega_r=0.0038$. (a) $g/\omega_r=0.0019$ (SC) (b) $g/\omega_r=0.1125$ (USC) (solid lines). The qubit gap is at $\varepsilon=0$ while the  photonic gaps are at  $\varepsilon=\pm\omega_r$. The color (dashed) lines are the energies $\pm\varepsilon/2$ ($\pm\varepsilon/2+\omega_r$) of the product states $\ket{\uparrow, 0}$ and $\ket{\downarrow, 0}$ ($\ket{\uparrow,1}$ and $\ket{\downarrow,1}$) in the absence of qubit-resonator coupling, i.e. for $\Delta, g=0$.}
	\label{fig:1}
\end{figure}

Before focusing on the  driven dynamics  we analyze the structure of  the  energy spectrum in the absence of driving,  i.e.  by replacing   $\varepsilon(t) \rightarrow \varepsilon= \varepsilon_0$ in the Hamiltonian of Eq.(\ref{eq:Hamil}). As we will show in the following sections this analysis  will be useful to interpret the interference LZS patterns once the driving is included.
In Fig.\ref{fig:1} we plot the three lowest energy levels as a function of $\varepsilon$, obtained after the numerical diagonalization of the Hamiltonian in the SC and the USC regimes, by choosing respectively  $g/\omega_r=0.0019$ (SC) and
$g/\omega_r=0.1125$ (USC). Although the qubit-resonator interaction mixes the states of the product basis in a non-trivial way, it is remarkable that for the parameters  considered  the eigenenergies can be  approximated by  $\pm\varepsilon/2+n\omega_r$, away from the avoided crossings. The associated eigenstates, spanned in the product basis, have weight  mainly on the states $\ket{\uparrow, n}$ and $\ket{\downarrow, n}$, respectively, with $n$ the number of photons in the resonator.

Due to  the  $\frac{\Delta}{2}\sigma_x$ term in the Hamiltonian $H_q$ of Eq.(\ref{eq:H_Rabi}), ``qubit gaps" of magnitude $\Delta$ open  at $\varepsilon=0$. Additionally, ``photonic gaps" $\Delta_n\approx2 g\sqrt{n+1}$ open at $\varepsilon=\pm\omega_r$ as a result of the qubit-resonator interaction \cite{blais_2004}. 

Our analysis goes beyond the dispersive regime, $g \ll |\varepsilon-\omega_r| \ll |\varepsilon + \omega_r|$, which is usually employed for non-demolition readout of the qubits in typical cQED proposals \cite{blais_2004}. In that scheme, the resonator experiences a frecuency shift that depends on the qubit state. Thus the state of the qubit can be read out indirectly by performing measurements on the resonator \cite{gu_2017}. 
However this technique relies on the use of an effective Hamiltonian and the range of parameters where the approximation is valid is restricted, while we report general results for arbitrary parameters.

\section{LZS interferometry}
\label{sec:II}

In order to study the LZS interferometry in cQED,
we  now include the  time dependent bias $\varepsilon(t)=\varepsilon_0+A\cos{\omega t}$.
To calculate the quantum dynamics of the full driven Rabi Hamiltonian, Eq.(\ref{eq:Hamil}),
we use the  Floquet formalism \cite{shirley_1965}, that allows for an exact treatment of time-periodic Hamiltonian $H(t)=H(t+\tau)$,
with $\tau=2\pi/\omega$ the period of the drive. 
In this formalism, the solutions of the Schr\"odinger equation $i\frac{d}{dt}\ket{\psi(t)}=H(t)\ket{\psi(t)}$ are expressed as $\ket{\psi_\alpha(t)}=e^{-i\varepsilon_\alpha t/}\ket{\alpha(t)}$, where the Floquet states $\ket{\alpha(t)}$ and corresponding quasienergies $\varepsilon_\alpha$ are obtained from the eigenvalue equation $\mathcal{H}\ket{\alpha(t)}=\varepsilon_\alpha\ket{\alpha(t)}$, being $\mathcal{H}=H(t)-i\partial_t$ the Floquet Hamiltoninan. The resulting Floquet states satisfy $\ket{\alpha(t)}=\ket{\alpha(t+\tau)}$\cite{son_2009,ferron_2010}.
We obtain numerically the Floquet states and quasienergies following the same procedure as in Ref.\onlinecite{ferron_2016}.

In  LZS interferometry,  when a quantum  system is   driven through an energy-level avoided crossing of magnitude $\tilde{\Delta}$ by a periodic signal of amplitude $A$ and frequency $\omega$, 
the resonance condition, for which the transfer of population is maximum,  depends on the velocity of passing through the avoided level crossing \cite{shevchenko_2010}.
Usually, the slow driving regime is  defined  for $A \omega < \tilde{\Delta}^2$, while the fast driving condition is  attained for  $A \omega \gg \tilde{\Delta}^2$.  In recent years, 
 the  specific features  of the associated LZS resonance patterns  have been studied and probed in driven qubits as both regimes have also been experimentally attained \cite{stehlik_2012,forster_2014, koski_2018, shevchenko_2018, gramajo_2020}.
 
\subsection{ The driven Jaynes-Cummings Hamiltonian: Photonic-LZS}
\label{subsec:IIA}

In this section  we analyze the Hamiltonian defined by Eq.(\ref{eq:H_Rabi}) under the assumption of $\Delta\ll\varepsilon_0 \sim \omega_r$. For relative small  driving amplitudes  such that
$A< \varepsilon_0 $,  we can neglect the term $\frac{\Delta}{2}\sigma_x$, as the system is always driven away from the qubit avoided crossing at $\varepsilon_0=0$ and the separation of the energy levels of the qubit $\omega_q=\sqrt{\varepsilon_0^2+\Delta^2}$ can be thus approximated by $\varepsilon_0$. In the present case and under the assumption $\{g, |\omega_q-\omega_r|\}\ll |\omega_q+\omega_r|$, the qubit-resonator interaction, $ g \sigma_y \left(a^\dagger+a\right)$, can be  replaced by $i g (\sigma _- a^\dagger - \sigma _+ a)$ as a result of a Rotating Wave Approximation (RWA)\cite{blais_2004, agarwal_2013,xie_2017}, where $\sigma_+$ and $\sigma_-$ are the raising and lowering operators for the qubit, respectively. This interaction conserves the number of excitations of the system qubit-resonator and makes  possible to solve separately the dynamics for each of the two-dimensional subspaces spanned by $\{\ket{\uparrow,n}, \ket{\downarrow,n+1}\}_{n\in \mathbb{N}}$ and the single state $\{\ket{\downarrow, 0}\}$. Under these assumptions the driven Jaynes-Cummings (DJC) Hamiltonian, defined in each of the mentioned subspaces\cite{blais_2004}, 

\begin{equation}
        \begin{split}
            H^{(n)}_{JC}(t)=(n+\frac{1}{2}) \omega_r \begin{pmatrix} 1 & 0 \\ 0 & 1 \end{pmatrix} + \frac{1}{2}\begin{pmatrix} \delta(t) & -i\Delta_n \\ i\Delta_n & -\delta(t) \end{pmatrix},
        \end{split}
    \label{eq:H_JC}
\end{equation}
is used to solve the system dynamics  instead of the original driven Rabi Hamiltonian, Eq.(\ref{eq:H_Rabi}).
As a consequence the system can be studied as  a  collection of (non-interacting) driven two level systems with a photon-number dependent gap $\Delta_n=2 g\sqrt{n+1}$, and energies globally shifted.
In Eq. (\ref{eq:H_JC}) we have defined  $\delta(t)\equiv \varepsilon(t)-\omega_r =\delta_0+A\cos{\omega t}$, with  $\delta_0= \varepsilon_0 - \omega_r$. 

In the following,  we  assume  that the system is prepared at the initial time $t=0$ in  the product state $\ket{\downarrow, n+1}$. After calculating the Floquet states and quasienergies of $H^{(n)}_{JC}(t)$, we  compute the time-averaged  probability (averaged over a period of the driving) of finding the system in the $\ket{\uparrow, n}$ state, $\overline{P}_{ \ket{\downarrow, n+1} \rightarrow  \ket{\uparrow, n}} =
\frac{1}{\tau} \int_{0}^{\tau}{ dt P_{ \ket{\downarrow, n+1} \rightarrow  \ket{\uparrow, n}} (t)} $.

\begin{figure}[H]

\centering
	
	\begin{overpic}[width=1\linewidth]{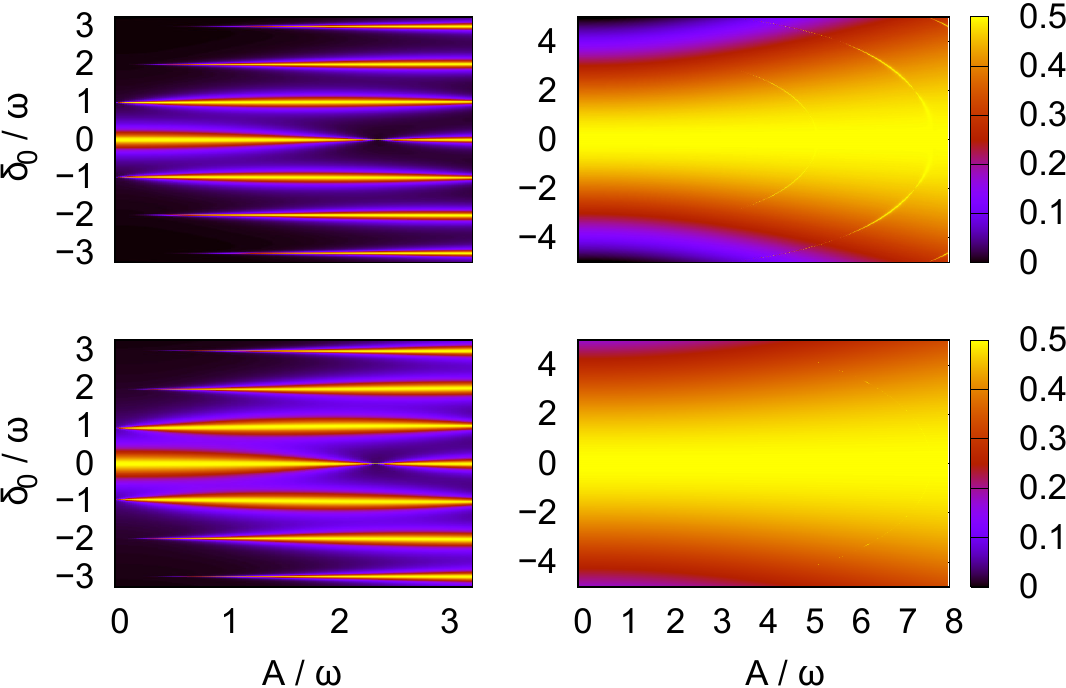}	
    \put(10,64.5){(a)}
    \put(53,64.5){(c)}
    \put(10,34){(b)}
    \put(53,34){(d)}
    \end{overpic}
    
\caption {Numerically obtained LZS interference patterns for the  DJC Hamiltonian, Eq.(\ref{eq:H_JC}). Plots of $\overline{P}_{\ket{\downarrow, n+1} \rightarrow \ket{\uparrow, n}}$ as a function of the  driving amplitude $A$ and dc bias $\delta_0$  in units of  $\omega $, for $g/\omega_r=0.0019$   and $n=3$ (a) ($n=10$ (b)) and $ g/\omega_r=0.1125$  and $n=0$ (c) ($n=1$ (d)).}
\label{fig:2}
\end{figure}
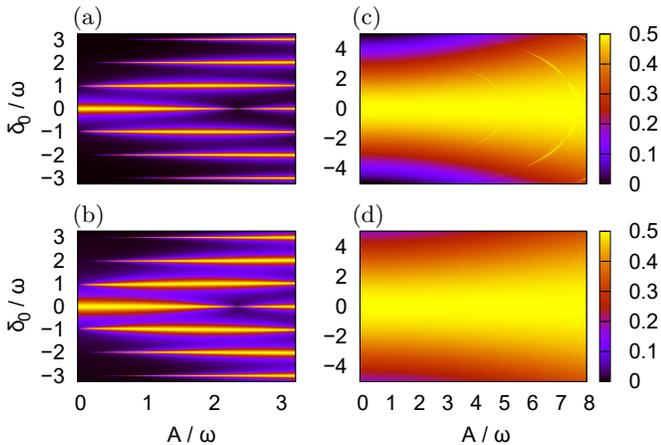

In Fig.\ref{fig:2} we show  the numerical results for the intensity plot of $\overline{P}_{\ket{\downarrow, n+1} \rightarrow  \ket{\uparrow, n}}$ as a function of the driving amplitude $A$ and dc detuning $\delta_0$, for the SC and USC regime and for two different values of $n$.  
For  $g/\omega_r=0.0019$ (see Fig.\ref{fig:2} (a) and (b)), as the photonic gap is small up to  values of  $n\gg 1$ ($ \Delta_n \sim 0.01 \sqrt {(n+1)}   \omega) $,  the system is in the fast driving regime. In this case, the resonance condition  is satisfied for  $\sqrt{\delta_0^2+\Delta_n^2}=m\omega$, which is the straighforward   generalization of the usual resonance condition obtained for  flux qubits in the fast driving regime \cite{oliver_2005,ferron2_2010, ferron_2016}.
For large detuning $\delta_0\gg \Delta_n$, the $m$-resonance condition becomes $\delta_0\approx m\omega$. 
Notice that the width of the resonance lobes  in   Fig.\ref{fig:2} (a) and (b) depends on the photon number $n$,  as  the magnitude of the avoided crossing is $\Delta_n=2 g\sqrt{n+1}$.
These  LZS  patterns can be qualitatively  described  by an analytical expression  for the average probability near the m-resonance, $\overline{P}^{RWA}_{\ket{\downarrow, n+1} \rightarrow \ket{\uparrow, n}} $, originally derived for driven flux qubits in the fast driving and within a Rotating Wave Approximation \cite{oliver_2005,ashhab_2007,shevchenko_2010,gramajo_2019}, and here trivially extended to analyze the DJC:

\begin{equation} 
    \overline{P}^{RWA}_{\ket{\downarrow, n+1} \rightarrow \ket{\uparrow, n}}=\frac{1}{2}\frac{[\Delta_n J_{-m}\left(\frac{A}{ \omega}\right)]^2}{[(\delta_0-m\omega)]^2+[\Delta_n J_{-m}\left(\frac{A}{ \omega}\right)]^2}. 
\label{eq:RWA_Formula}
\end{equation}
This  equation  shows  Lorentzian-shaped  resonances with a  maximum probability value of $\frac{1}{2}$  at $\delta_0=m\omega$ and width $|\Delta_n J_{-m}\left(\frac{A}{ \omega}\right)|$, being $J_{m}(x)$ the $m^{th}$ order Bessel function of the first kind.  In particular, at the zeros of $J_{-m}(x)$ is $\overline{P}^{RWA}_{\ket{\downarrow, n+1} \rightarrow \ket{\uparrow, n}}=0$, a phenomenon known as \textit{coherent destruction of tunneling}\cite{grifoni_1998}.
 
\begin{figure}[H]
\centering
	\begin{overpic}[width=1\linewidth]{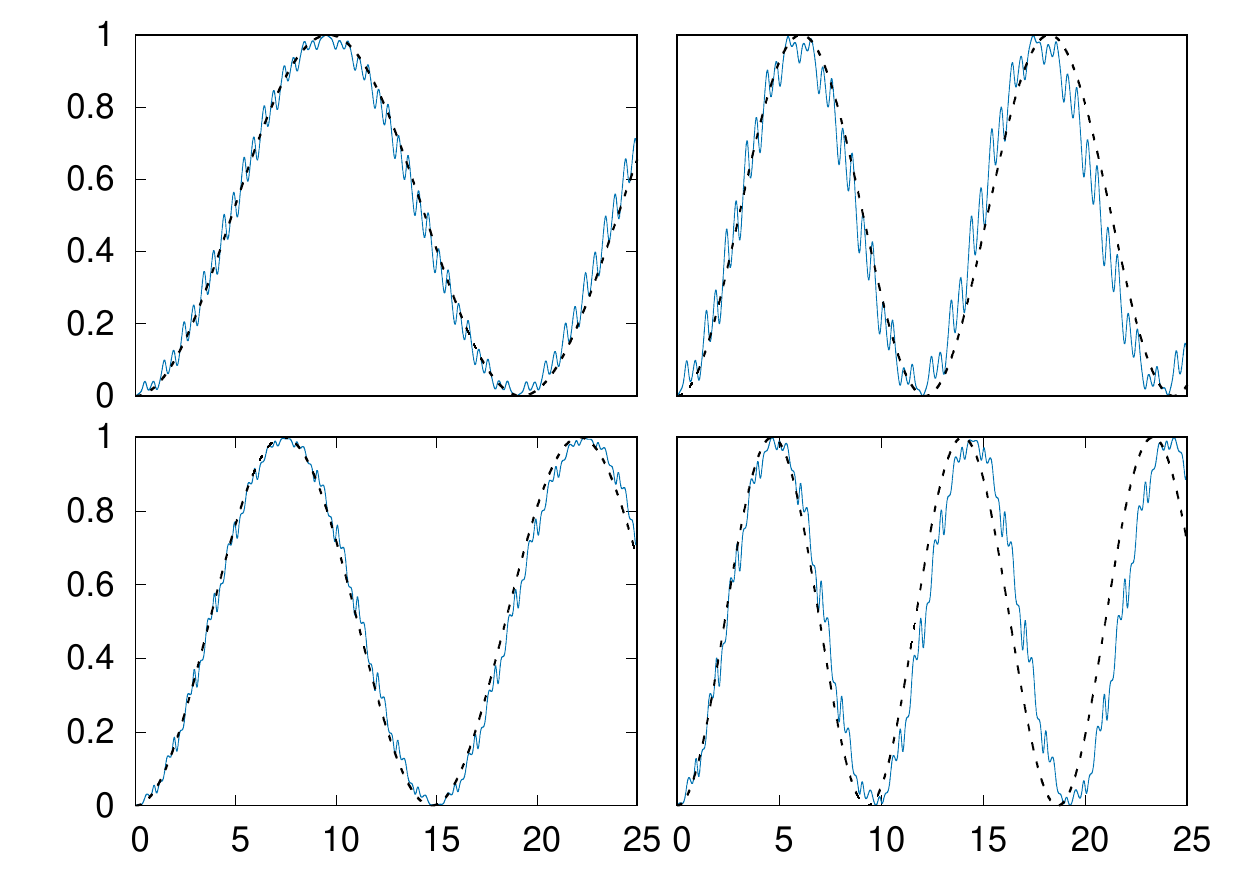}

    \put(11.5,63.5){(a)}
    \put(54.5,63.5){(b)}
    \put(11.5,31.5){(c)}
    \put(54.5,31.5){(d)}
    
    \put(30,-2){$t [\tau]$}
    \put(73,-2){$t [\tau]$}
    \put(0,8){\rotatebox{90}{$P_{\ket{\downarrow, n+1}\rightarrow \ket{\uparrow, n}}(t)$}}
    \put(0,40){\rotatebox{90}{$P_{\ket{\downarrow, n+1}\rightarrow \ket{\uparrow, n}}(t)$}}
    
    \end{overpic}
	\caption {Instantaneous transition probability  $P_{ \ket{\downarrow, n+1} \rightarrow  \ket{\uparrow, n}} (t)$
	for driving amplitude $A/\omega=3$ and coupling $g/\omega_r=0.0019$.
	Top panel: photon number $n=3$ (a) and $n=10$ (b) for  resonance $\delta_0/\omega=0$. Lower panels: photon number $n=3$ (c) and $n=10$ (d) for resonance $\delta_0/\omega=1 $. The fast oscillations exhibited in the
	numerical results (solid line), are not captured by the RWA (dashed line).}
	\label{fig:3}
	\end{figure}
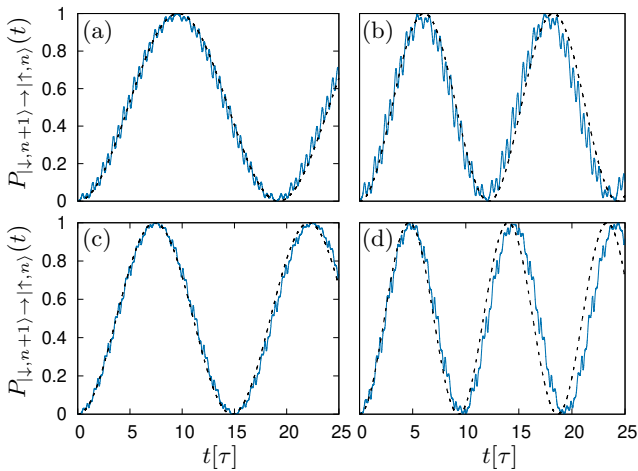

Despite of the fact that the  shape and positions of the resonances in Fig.\ref{fig:2} (a) and (b) are captured by Eq.(\ref{eq:RWA_Formula}),  the  instantaneous transition probability  $P_{ \ket{\downarrow, n+1} \rightarrow  \ket{\uparrow, n}} (t)$ depicted in  Fig.(\ref{fig:3}) exhibits   fast oscillations  which are not captured by
the RWA.  Notice  that the period of these fast oscillations  depends  on the value $n$, which changes the effective gap  $\Delta_n$ and therefore the structure of the instantaneous transition probability.

Upon increasing $g$ or increasing $n$ the photonic gap becomes much larger than the driving frequency $\omega$ and the LZS interferometry patterns correspond to the  slow driving regime. In Fig.\ref{fig:2} (c) and (d) we show the USC  case for $g/\omega_r=0.1125$, where already for $n=0$ is  $\Delta_n >  \omega$.
In this case the resonances describe arcs around the point $A=0$, $\delta_0=0$ \cite{sillanpaa_2006,shevchenko_2010, koski_2018}. Notice that for $n=1$  and for  the
values of $A$ considered, the complete adiabatic regime is attained.

\subsection{The driven Rabi Hamiltonian: Combined Photonic-LZS + Qubit-LZS}
\label{subsec:IIB}

Away from the regime analyzed in the previous section, the full driven  Rabi (DR) Hamiltonian  Eq.(\ref{eq:H_Rabi}) has to be solved.  The effect of the counter rotating terms become important either because of ultra-strong coupling or because of extremely large detuning, $\{g, |\omega_q-\omega_r|\}\sim |\omega_q+\omega_r|$. Under these conditions, the RWA  that gave place to the DJC Hamiltonian , Eq.(\ref{eq:H_JC}), breaks down \cite{niem_2010, agarwal_2013, forn_2019}.

In the following, we analize the LZS interferometry patterns that emerge  for the DR Hamiltonian, i.e. when the time dependence $\varepsilon(t)=\varepsilon_0 + A\cos{\omega t}$ is taken into account.
 As we will show, the different  avoided crossings present in the spectrum of Fig.\ref{fig:1} will produce a richer and more complex structure in the LZS  patterns in comparison to  those obtained for the DJC effective two-level system. 
 
In order to make the calculations numerically affordable,  and without loss of generality in our analysis, we  consider up to $n=3$ photons and calculate the time-averaged probability, $\overline{P}_{ \ket{\uparrow}}$, of measuring the qubit in the state $\ket{\uparrow}$ regardless the number of  photons in the resonator, for the initial condition $\ket{\downarrow, 0}$, as a function of the
driving amplitude $A$ and the dc  bias $\varepsilon_0$.

Different  resonance conditions - dependent on  the driving amplitude $A$, the magnitude  of the different gaps and their relative position with respect to the dc bias $\varepsilon_0$- contribute  to the interference patterns.
A gap can mediate a  LZS transition only if it is reached by the driving range $(\varepsilon_0-A,\varepsilon_0+A)$. Therefore, for a given value of $\varepsilon_0$ and for increasing amplitudes  starting at $A=0$, different avoided crossings can be accessed \cite{oliver_2005,berns_2008}. 

\begin{figure}[H]
	\begin{subfigure}[ht!]{0.5\textwidth} 	
	\centering
		\begin{overpic}[width=0.85\textwidth]{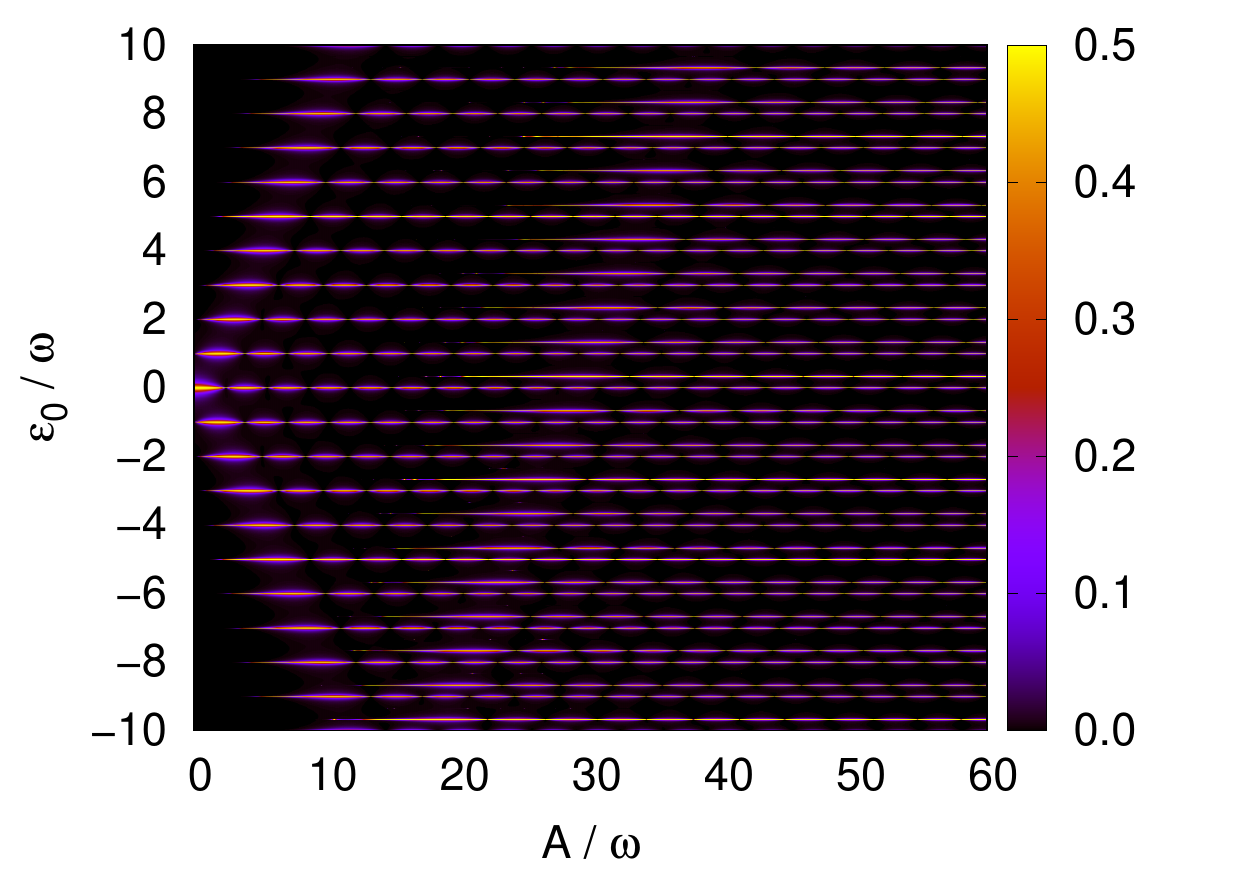}
       \put(0,62){(a)}
       \put(49,12.5){\includegraphics[width=0.87in]{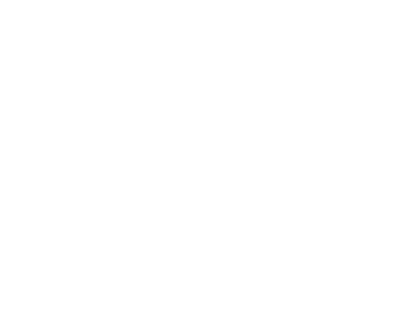}}
        \put(49.5,14){\includegraphics[width=0.9in]{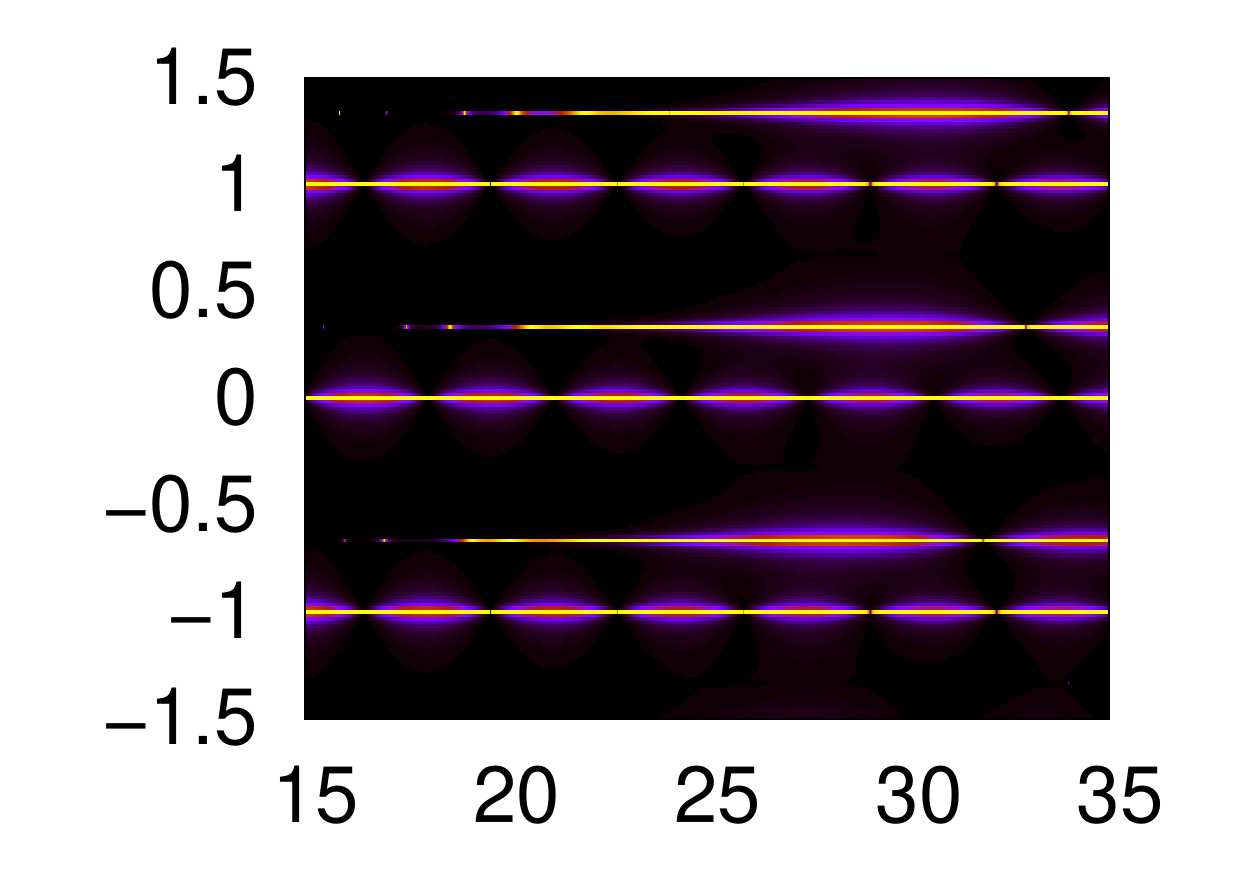}}
        \put(64,13){\includegraphics[width=0.15in]{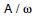}}

        \put(49.5,22){{\includegraphics[width=0.08in]{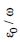}}}        
    \end{overpic}
        \end{subfigure}
	\begin{subfigure}[ht!]{0.5\textwidth} 
	\centering
		\begin{overpic}[width=0.85\textwidth]{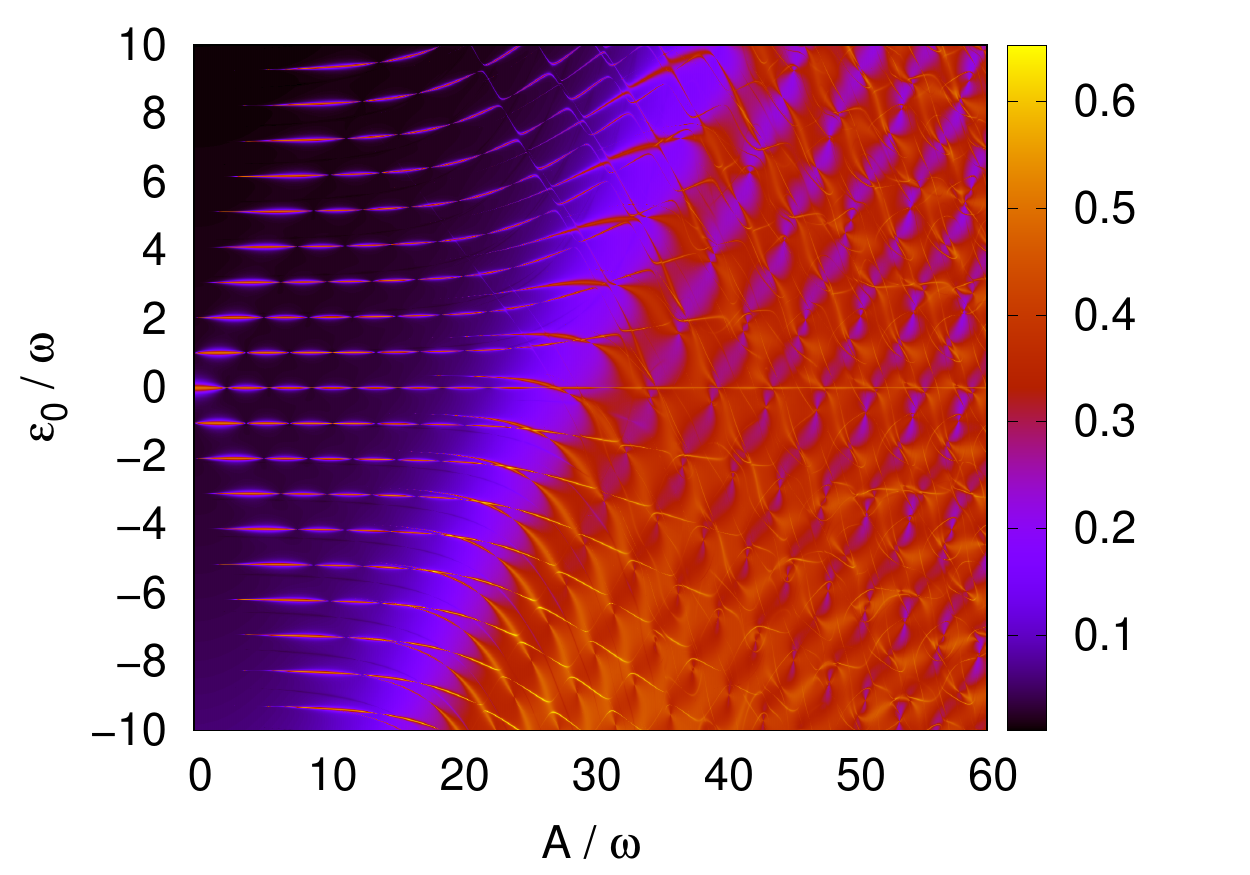}
       \put(0,62){(b)}
        \put(49,12.5){\includegraphics[width=0.87in]{Imagenes_nuevas/FIG_4_blank.JPG}}
        \put(49.5,14){\includegraphics[width=0.9in]{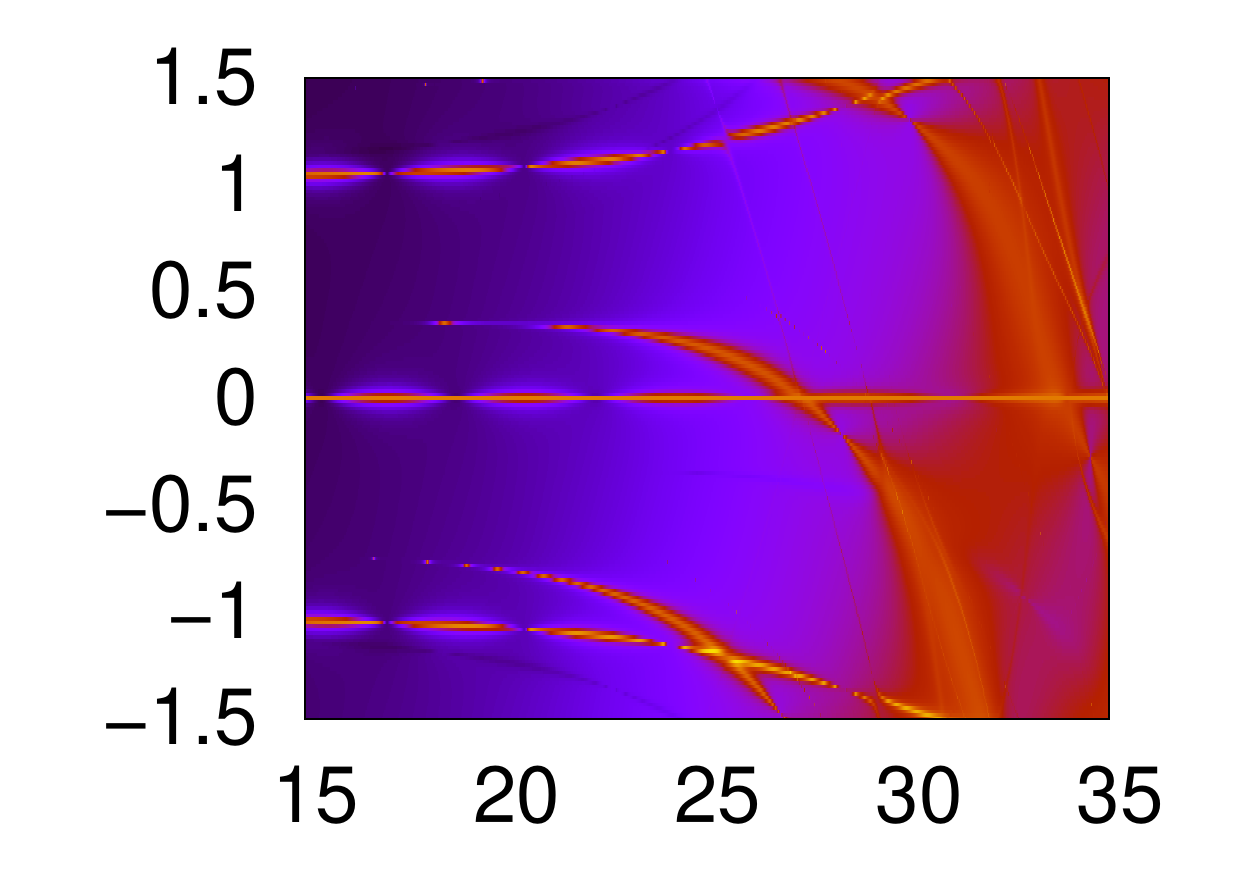}}
        \put(64,13){\includegraphics[width=0.15in]{Imagenes_nuevas/FIG_4_eje_x.JPG}}

        \put(49.5,22){{\includegraphics[width=0.08in]{Imagenes_nuevas/FIG_4_eje_y.JPG}}}        

        \end{overpic}
	\end{subfigure}
	\caption{Intensity plots of the LZS interference patterns for the driven Rabi Hamiltonian Eq.(\ref{eq:H_Rabi}). Plots of $\overline{P}_{ \ket{\uparrow}}$ as a function of the driving parameters $A$ and $\varepsilon_0$, for $g/\omega_r=0.0019$ (a) and $g/\omega_r=0.1125$ (b). The calculations were performed for $\omega/\omega_r=0.0375$ and $\Delta/\omega_r=0.0038$.
	The insets in both panels show the resonances patterns in more detail.}
	\label{fig:4}
\end{figure}

We start by analyzing the intensity plot of $\overline{P}_{ \ket{\uparrow}}$  for the SC regime  ($g/\omega_r=0.0019$),
in terms of  $A$ and $\varepsilon_0$.
Fig.\ref{fig:4}(a) exhibits resonances  characteristic of  the fast driving  \cite{shevchenko_2010, ferron_2010}. To understand its structure, it is instructive to  focus on the three lowest energy levels of the spectrum shown in Fig.\ref{fig:1}(a). The qubit central gap separating the first and second levels is involved in the  $\ket{\downarrow, 0} \leftrightarrow \ket{\uparrow, 0}$ transitions, with the associated qubit resonance condition $\varepsilon_0=m\omega$. The photonic gap at $\varepsilon_0=-\omega_r$ mediates $\ket{\downarrow, 0} \leftrightarrow \ket{\uparrow,1}$ transitions, with a resonance condition given by $\varepsilon_0=-\omega_r + m\omega$. 
On the other hand, the photonic gap at $\varepsilon_0=\omega_r$  favors $\ket{\uparrow, 0} \leftrightarrow \ket{\downarrow,1}$ transitions, with a resonance condition $\varepsilon_0=\omega_r + m\omega$. 
The LZS interference associated with this later gap does not contribute to  $\overline{P}_{ \ket{\uparrow}}$ since in the present case we start with the initial condition  $\ket{\downarrow, 0}$.
The different qubit and photonic resonances conditions  are  thus organized along  (shifted) horizontal lines, giving place to 
the pattern exhibited in  Fig.\ref{fig:4}(a), where the ``qubit-LZS'' interference pattern and the ``photonic-LZS'' interference pattern are combined.

\begin{figure}[H]
\centering
	\begin{overpic}[width=0.8\linewidth]{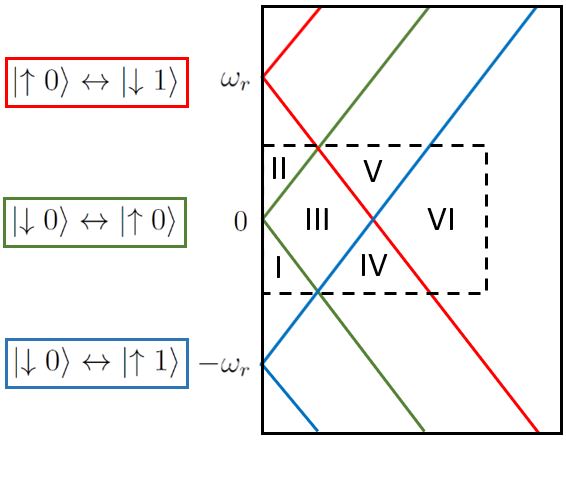}	
    \put(40,80){\large$\varepsilon_0$}
    \put(95,0){\large$A$}
    \end{overpic}
    \caption {(solid lines) Scheme of the boundaries of the regions where the LZS resonances mediated by the different avoided crossings occur in plots of $\overline{P}_{\ket{\uparrow}}$, as a function of the driving parameters $\varepsilon_0$ and $A$. States involved in the transitions are indicated. A rectangle in dashed lines shows the range of the parameters considered along this work.}
\label{fig:5}
\end{figure}

In Fig.\ref{fig:5} we show a scheme depicting the boundaries of the regions where resonances associated to the qubit and photonic avoided crossings  occur in   $\overline{P}_{ \ket{\uparrow}}$,  as a function of the dc bias $\varepsilon_0$ and amplitude $A$. The dashed-line rectangle delimits the region of parameters considered in this work. We label six different regions according to the resonances that appear in the patterns. Following the description of Fig.\ref{fig:4} (a), we see that the  regions I and II present no resonances. In region III there is a pure qubit-LZS pattern with only the resonances mediated by the qubit gap at $\varepsilon_0=0$.  In region IV coexist the combined qubit-LZS pattern and the photonic-LZS pattern associated to  the gap  at $\varepsilon_0=-\omega_r$.  Analogously, in region V coexist the  qubit-LZS pattern and the photonic-LZS pattern due to  the gap  at $\varepsilon_0=\omega_r$ (seen when the initial condition has components in  $\ket{\uparrow, 0}$). Finally, region VI contains  the combined qubit-LZS pattern and both  photonic-LZS patterns.

Fig.\ref{fig:4}(b) exhibits the case of ultra strong coupling with
$g/\omega_r=0.1125$. We observe a different structure of resonances,  as a  consequence of the enlargement of the photonic gaps located at $\varepsilon_0=\pm\omega_r$.
The resonances associated to the gap at $\varepsilon_0=-\omega_r$ form arcs around the point $A=0$, $\varepsilon_0=-\omega_r=-26.667\omega$ (partially  observed in the inset of Fig.\ref{fig:4}(b)), which as we have already mentioned, are expected for the slow driving regime \cite{shevchenko_2018,koski_2018}. 
However in our work  the slow driving regime is attained due to the increase in the value of $\Delta_n$, instead of reducing the  driving frequency $\omega$.  Additionally, the lobe-shaped resonances associated  to  the central qubit  gap $\Delta$ are distorted for large amplitudes $A$ in comparison to the SC ($g/\omega_r=0.0019$) case analyzed in Fig.\ref{fig:4}(a).
Notice that  the maximum value of $\overline{P}_{ \ket{\uparrow}}$ in  Fig.\ref{fig:4}(b) is larger than $1/2$ due to the superposition of  different resonances and in contrast to Fig.\ref{fig:4}(a), where resonances are isolated
and $\overline{P}_{ \ket{\uparrow}} \le 1/2$.

The qualitative differences  exhibited between  Fig.\ref{fig:4} (a) and (b) 
are thus mainly related to the increase in the size of the photonic 
gaps $\Delta_n$ as  $g/\omega_r$  increases from the SC to the USC regime. Notice that   the qubit-LZS interference patterns correspond always to the fast driving regime as we  consider $\omega = 10 \Delta$ along this work.

\section{Dissipative  effects}
\label{sec:III}

Experimentally, the system is affected by the electromagnetic environment that introduces decoherence and relaxation, affecting the quantum phase of the superposition states, and/or causing spontaneous decay of
the population.
Accordingly, any realistic approach to model and study the dynamics of the quantum system must take the coupling to the environment into account. To include dissipative effects, we  consider that the resonator is weakly coupled to a thermal reservoir modeled as an infinite set of non-interacting harmonic oscillators \cite{breuer_petruccione_2006}. This assumption is justified in typical cQED architectures in which the superconducting qubit is fabricated inside a transmission line resonator \cite{blais_2004, yan_2018}.  Thus, when the qubit and the resonator are  off-resonant ($\omega_q\neq\omega_r$), the resonator effectively acts as a filter of the environmental noise for the qubit. As a result, the  qubit coherence times are enhanced because it is only indirectly coupled to the external noise sources through  the transmission line.

The general theoretical  approach to study open systems is to consider a  total  (system plus bath) Hamiltonian  given by 
\begin{equation}
H(t)=H_S(t)+H_B+H_{SB}\;,
\label{eq:H_universe}
\end{equation}
where $H_S$ is the system Hamiltonian and 
\begin{equation}
        \begin{split}
            &H_B=\sum_\nu \omega_\nu b_\nu^\dagger b_\nu, \\ 
            &H_{SB}=(a+a^\dagger)\sum_\nu c_\nu (b_\nu+b_\nu^\dagger)+(a+a^\dagger)^2\sum_\nu \frac{c_\nu^2}{\omega_\nu},
       \end{split}
    \label{eq:H_B_H_SB}
\end{equation}
are the terms for the bath and the system-bath interaction. In the cQED architecture here considered, the  bath oscillators have frequencies $\omega_\nu$, with $b_\nu^\dagger$ ($b_\nu$) the  creation (anhilation) operators, and are linearly coupled to the resonator operator $(a+a^\dagger)$, with coupling strength $c_\nu$. Following the usual approach,
the bath is characterized by a continuous distribution of modes with an  ohmic spectral density $J(\omega)= \kappa\omega e^{-\omega/\omega_D}$, with damping constant $\kappa$ and cutoff frequency $\omega_D$. 

The time evolution of the reduced density matrix is computed after expanding $\rho(t)$ in terms of  the time-periodic  Floquet basis, $\rho_{\alpha\beta}(t)=\langle {\alpha}(t)|\rho(t)|{\beta}(t)\rangle$, and performing the Born (weak coupling) and Markov (fast relaxation) approximations. The resulting
 Floquet-Born-Markov (FBM) master equation \cite{kohler_1998} is solved numerically,  and with it we  compute $\overline{P}_{\ket{\uparrow}}(t)$. For details on these calculations  we refer the reader to  references \onlinecite{hausinger_2010, ferron_2016, gramajo_2019}.

We  consider  the system as composed by the qubit and the resonator and described by the DR Hamiltonian Eq. (\ref{eq:H_Rabi}). Following the Born approximation, the
resonator is  assumed as weakly coupled to the ohmic thermal reservoir. Thus, for the numerical results -and consistent with typical experimental parameters- we  take $\kappa=0.001$, corresponding to weak dissipation, and a large cutoff  frequency $\omega_D=12.5\omega_r$. The bath temperature  is $T=0.0175\,\omega_r/k_B$ ($\sim20\,mK$).

\begin{figure}[H]
	\begin{subfigure}[ht!]{0.5\textwidth} 	
		\centering
		\begin{overpic}[width=0.85\textwidth]{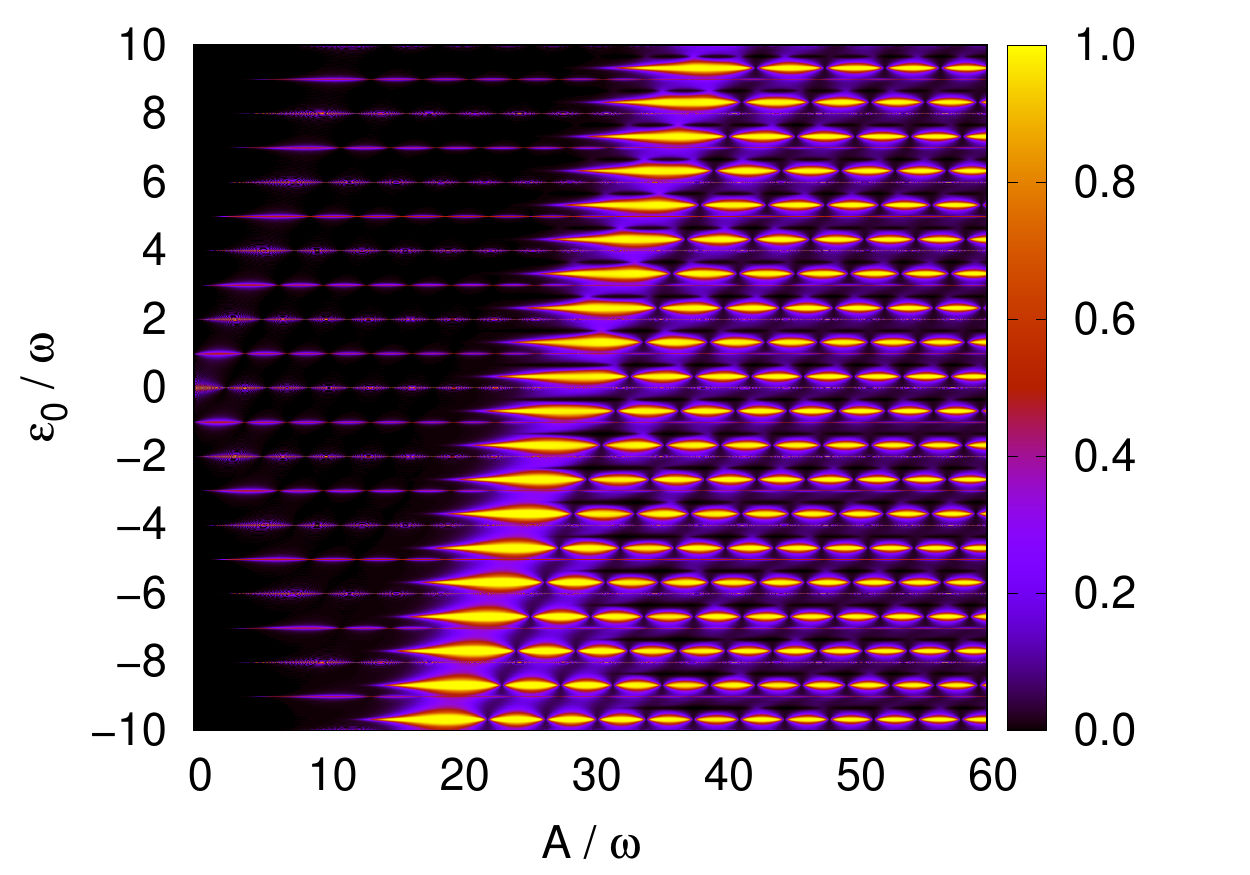}
			\put(0,62){(a)}
		\end{overpic}
	\end{subfigure}
	\begin{subfigure}[ht!]{0.5\textwidth} 
		\centering
		\begin{overpic}[width=0.85\textwidth]{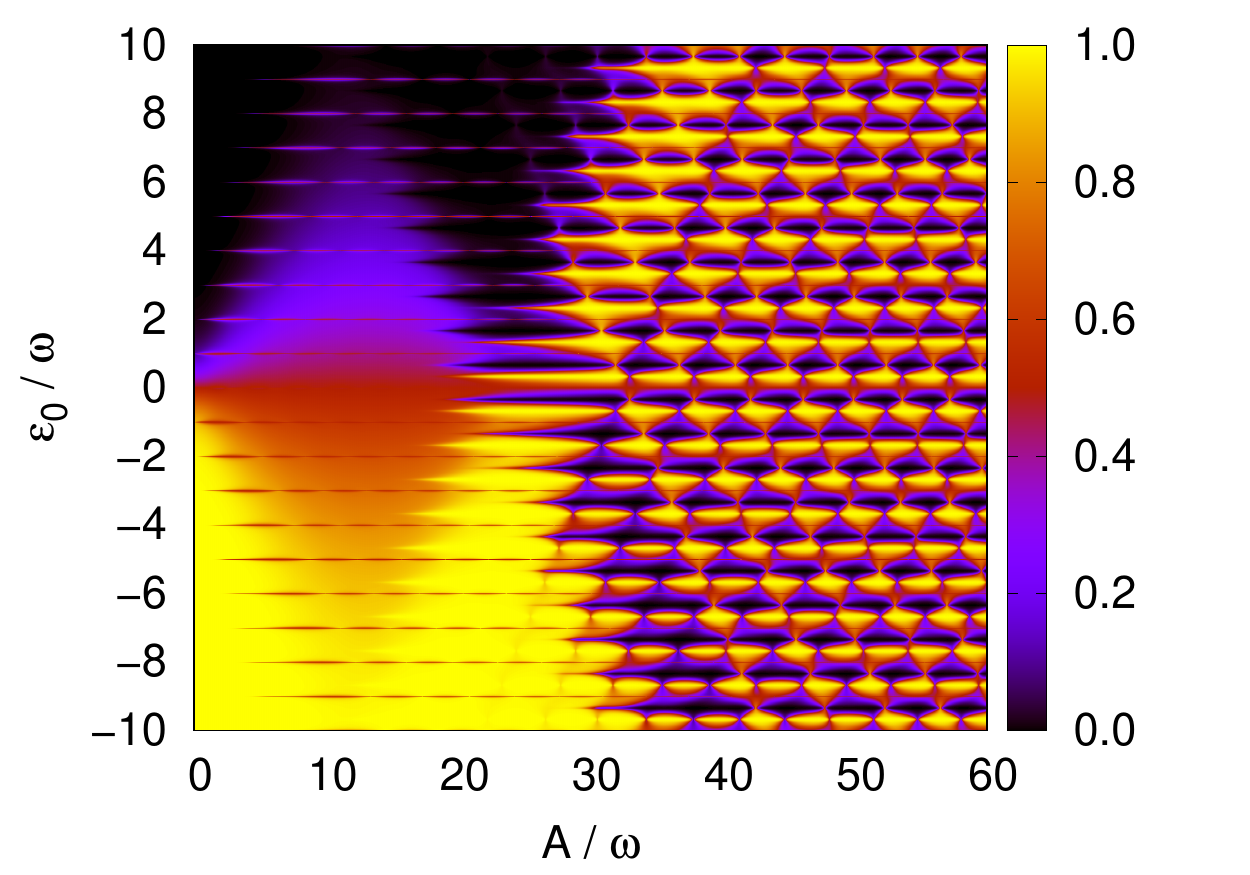}
			\put(0,62){(b)}
		\end{overpic}
	\end{subfigure}
	\caption{LZS interference patterns for the Rabi Hamiltonian in the SC Regime ($g/\omega_r=0.0019$)  considering effects of dissipation. Plots of $\overline{P}_{\ket{\uparrow}}(t)$ as a function of the driving parameters $A$ and $\varepsilon_0$, at finite time $t=1000\tau$ (a) and in the asymptotic regime $t=\infty$ (b). The calculations were performed for $\omega/\omega_r=0.0375$, $\Delta/\omega_r=0.0038$, $T=0.0175\,\omega_r/k_B$ and $\kappa=0.001$ (see text for details).}
	\label{fig:6}
\end{figure}

In Fig.\ref{fig:6} we plot $\overline{P}_{\ket{\uparrow}}(t)$ (for the initial condition $\ket{\downarrow, 0}$) as a function of $A$ and $\varepsilon_0$ in the strong coupling case, for $g/\omega_r=0.0019$. 

\begin{figure}[ht]
	\begin{subfigure}[ht!]{0.5\textwidth} 	
		\centering
		\begin{overpic}[width=0.85\textwidth]{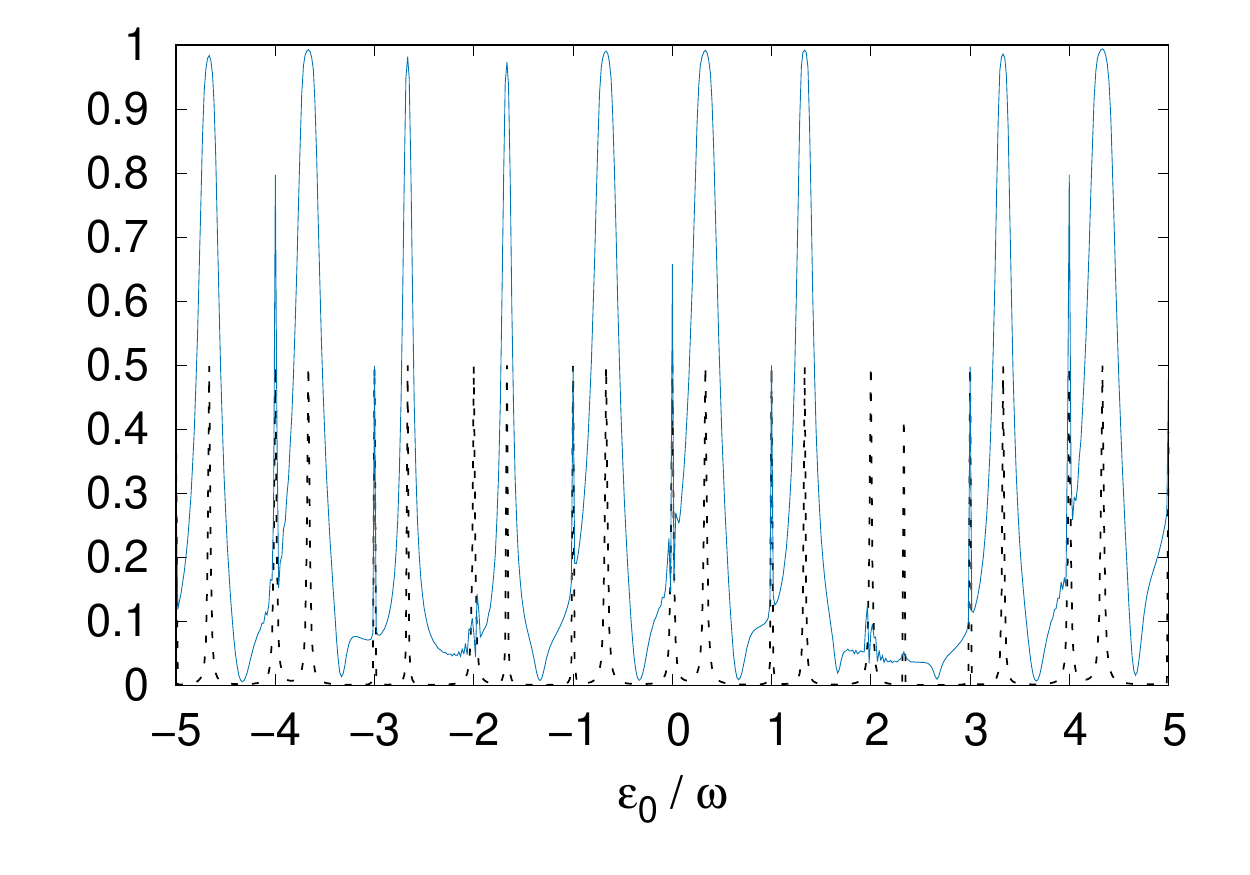}
			\put(0,62){(a)}
			\put(0,35){\rotatebox{90}{$\overline{P}_{\ket{\uparrow}}$}}
		\end{overpic}
	\end{subfigure}
	\begin{subfigure}[ht!]{0.5\textwidth} 
		\centering
		\begin{overpic}[width=0.85\textwidth]{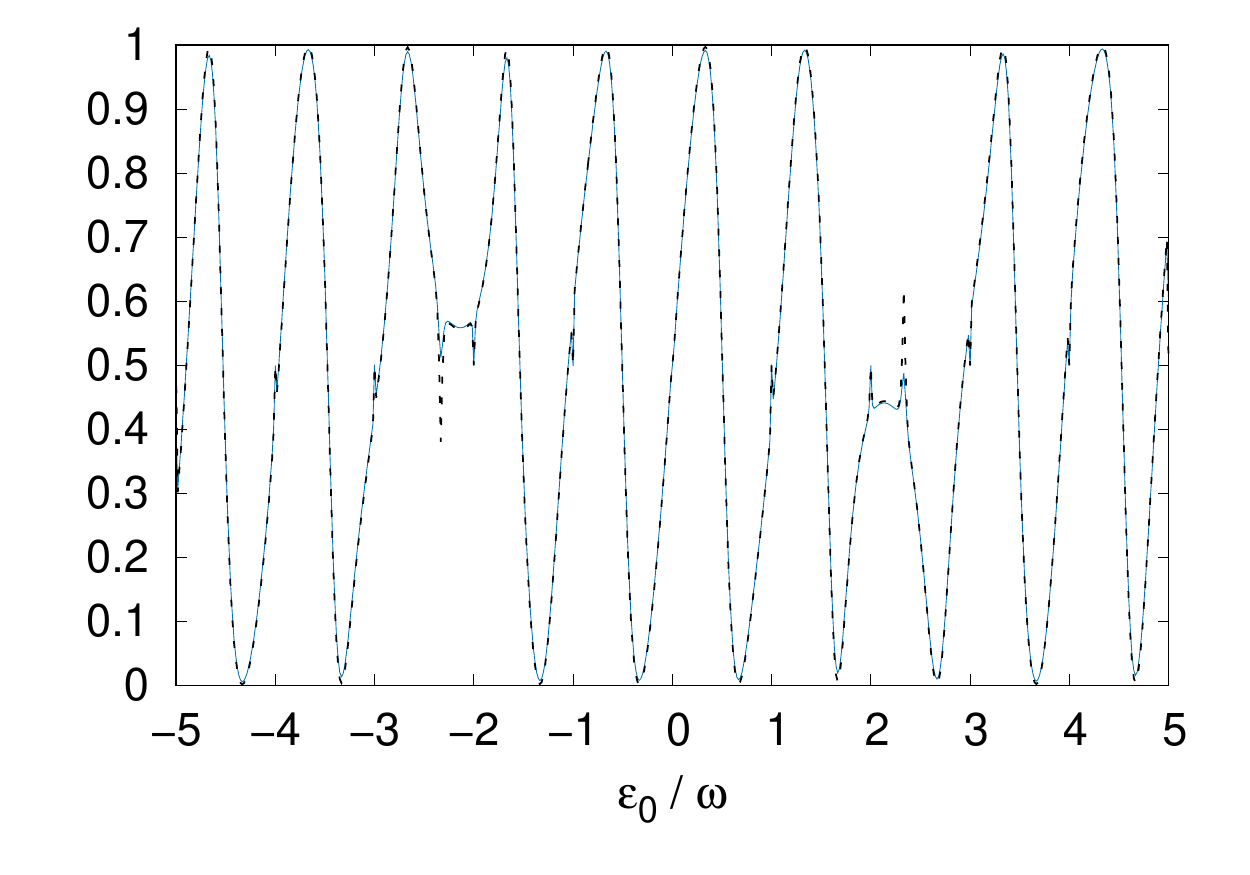}
			\put(0,62){(b)}
			\put(0,35){\rotatebox{90}{$\overline{P}_{\ket{\uparrow}}$}}
		\end{overpic}
	\end{subfigure}
	\caption{(a) Avareged transition probability for the closed system $\overline{P}_{\ket{\uparrow}}$ (dashed line) and considering effects of dissipation at finite time $\overline{P}_{\ket{\uparrow}}(t\rightarrow1000\tau)$ (solid line) as a function of $\varepsilon_0$ for $g/\omega_r=0.0019$ and $A=35\,\omega$.
		(b) Avareged transition probability in the asymptotic regime $\overline{P}_{\ket{\uparrow}}(t\rightarrow\infty)$ as a function of $\varepsilon_0$ for $g/\omega_r=0.0019$ and $A=35\,\omega$. Numerical results when the system is taken as qubit plus resonator (DR Hamiltonian) coupled to an ohmic bath (solid line) and when we regard the driven qubit coupled to a structured bath (dashed lines).}
	\label{fig:7}
\end{figure}

First, we describe in Fig.\ref{fig:6}(a) the results for finite time $t=1000\tau$. At this finite time, the spectroscopic pattern  still reflects the effect of the initial condition $\ket{\downarrow, 0}$, and resembles  the one  obtained for the unitary evolution, Fig.\ref{fig:4}(a).
However,  when we compare both patterns  two main  differences emerge in Fig.\ref{fig:6}(a) : i) the resonances associated to the photonic gaps at $\varepsilon=-\omega_r$ are broadened 
and ii) the probability $\overline{P}_{\ket{\uparrow}}(t)$ takes values close to 1. These features  are better seen in Fig.\ref{fig:7}(a), where we plot $\overline{P}_{\ket{\uparrow}}$ as a function of $\varepsilon_0$ for  $A/\omega=35$. In the case of the unitary evolution, we observe narrow peaks corresponding to the qubit resonances at $\varepsilon_0=n\omega$ and  peaks corresponding to the photonic resonances at $\varepsilon_0=-\omega_r + m\omega$. After  adding dissipation, the qubit resonances stay narrow while the photonic resonances broaden and  take values close to $\overline{P}_{\ket{\uparrow}}=1$. In this later case at the photonic resonance there is a transfer of population among the $\ket{\downarrow, 0} \leftrightarrow \ket{\uparrow,1}$  states followed by decay transitions $\ket{\uparrow,1}\rightarrow \ket{\uparrow, 0}$ induced by the dissipative coupling with the bath. In this way, when the initial condition is $\ket{\downarrow, 0}$ the  effect of the ac drive and  the dissipation  is to  continuously pump population
from  $\ket{\downarrow, 0}$ to  $\ket{\uparrow, 0}$, leading to $\overline{P}_{\ket{\uparrow}}\approx 1$. 
For $\varepsilon_0>0$ the ground state has a principal weight on state $\ket{\downarrow, 0}$, and therefore 
$\overline{P}_{\ket{\uparrow}}\approx 1$ corresponds to population inversion, and the resonance can be interpreted as a blue sideband resonance \cite{hausinger_2010,ferron_2016}.
On the other hand, for $\varepsilon_0<0$ the ground state has a principal weight on state $\ket{\uparrow, 0}$, and therefore 
$\overline{P}_{\ket{\uparrow}}\approx 1$ corresponds to full cooling into the ground state, as in a red sideband resonance \cite{hausinger_2010,ferron_2016}.

Effects of the bath are much more notorious in the steady state regime for $t\rightarrow\infty$  (Fig.\ref{fig:6}(b)) where the characteristic diamond-like spectroscopy patterns are easily identified. The steady state is independent of  the initial condition, and thus the asymptotic pattern combines the effect of the photonic gaps at $\varepsilon_0=-\omega_r$ and at $\varepsilon_0=\omega_r$. 
This can be understood in terms of the energy spectrum shown in Fig. \ref{fig:1} and the different regions defined in Fig.\ref{fig:5}. 

In region I (II) of Fig.\ref{fig:5} where resonances are absent, only the dissipative contribution is present, and thus  $\overline{P}_{\ket{\uparrow}}(t\rightarrow\infty) \sim 1(0)$, respectively. 
Region III, corresponding to the first diamond,  presents a qubit-LZS pattern with narrow lobe-shaped qubit resonances at $\varepsilon_0=n\omega$ and a background structure due to  the relaxation processes.
Regions IV and V correspond to the intermediate sector between diamonds, for $\varepsilon_0<0$ and $\varepsilon_0>0$, respectively.
In the second diamond sector (Region VI), the combined effect of the resonances associated to the qubit gap  and the two photonic gaps contribute to the LZS transitions.  The photonic resonances at $\varepsilon_0=-\omega_r + m\omega$ give  maxima with $\overline{P}_{\ket{\uparrow}}\approx 1$, due to the LZS transition plus decay mechanism described above.
On the other hand, the photonic resonances at $\varepsilon_0=\omega_r + m\omega$ give  $\overline{P}_{\ket{\uparrow}}\approx 0$. In this case, at the photonic resonance there is a transfer of population between  $\ket{\uparrow, 0} \leftrightarrow \ket{\downarrow,1}$  states plus dissipative decay transitions $\ket{\downarrow,1}\rightarrow \ket{\downarrow, 0}$, leading to $\overline{P}_{\ket{\uparrow}}\approx 0$. In Fig.\ref{fig:7}(b) we see clearly the two types of photonic resonances with  alternating broad peaks with $\overline{P}_{\ket{\uparrow}}\approx 1$ and broad dips with $\overline{P}_{\ket{\uparrow}}\approx 0$, in a plot of the dependence of $\overline{P}_{\ket{\uparrow}}$ with  $\varepsilon_0$ for $A=35\,\omega$. It is also possible to notice the narrow peaks corresponding to the qubit resonances at $\varepsilon_0=n\omega$.

For the analysis of Fig.\ref{fig:6}(b) it is also worthwhile to mention that for  $g\ll\omega_{q/r}$,  a plausible  assumption  is to consider  the quantum system solely as the qubit. In this approach,  the transmission line resonator is taken as a part of the environment seen by the qubit and it is possible to map  the composite (resonator-bath)  reservoir to a ''structured bath" of non-interacting harmonic oscillators with an effective spectral density 
\begin{equation}
J_{eff}(\omega)= \frac{16\kappa g^2 \omega_r^2\omega}{(\omega_r^2-\omega^2)^2+(\kappa\omega_r\omega)^2},
\label{eq:J_efectiva_estructurado}
\end{equation}
that behaves as ohmic at low frequencies and presents a Lorentzian peak at $\omega=\omega_r$ 
\cite{garg_1985,goorden_2004,ferron_2016}. The qubit-structured bath coupling Hamiltonian  is of the form
$H_{qB}\propto \sigma_y X$, with $X$ a coordinate of the structured bath.
In Fig.\ref{fig:7}(b) we see a good agreement between the $\overline{P}_{\ket{\uparrow}}$ obtained from the qubit coupled through $\sigma_y$ to an structured bath and the corresponding results of Fig.\ref{fig:6}(b) for the DR Hamiltonian.
Within the ''qubit + structured bath" scenario one can interpret the qubit LZS pattern of  Region III
as the LZS pattern of a qubit  transversely coupled  to  a bath, studied  in Ref. \onlinecite{gramajo_2019}, which is characterized by narrow resonance peaks and a background off-resonance population.

\begin{figure}[t]
	\begin{subfigure}[ht!]{0.5\textwidth} 	
		\centering
		\begin{overpic}[width=0.85\textwidth]{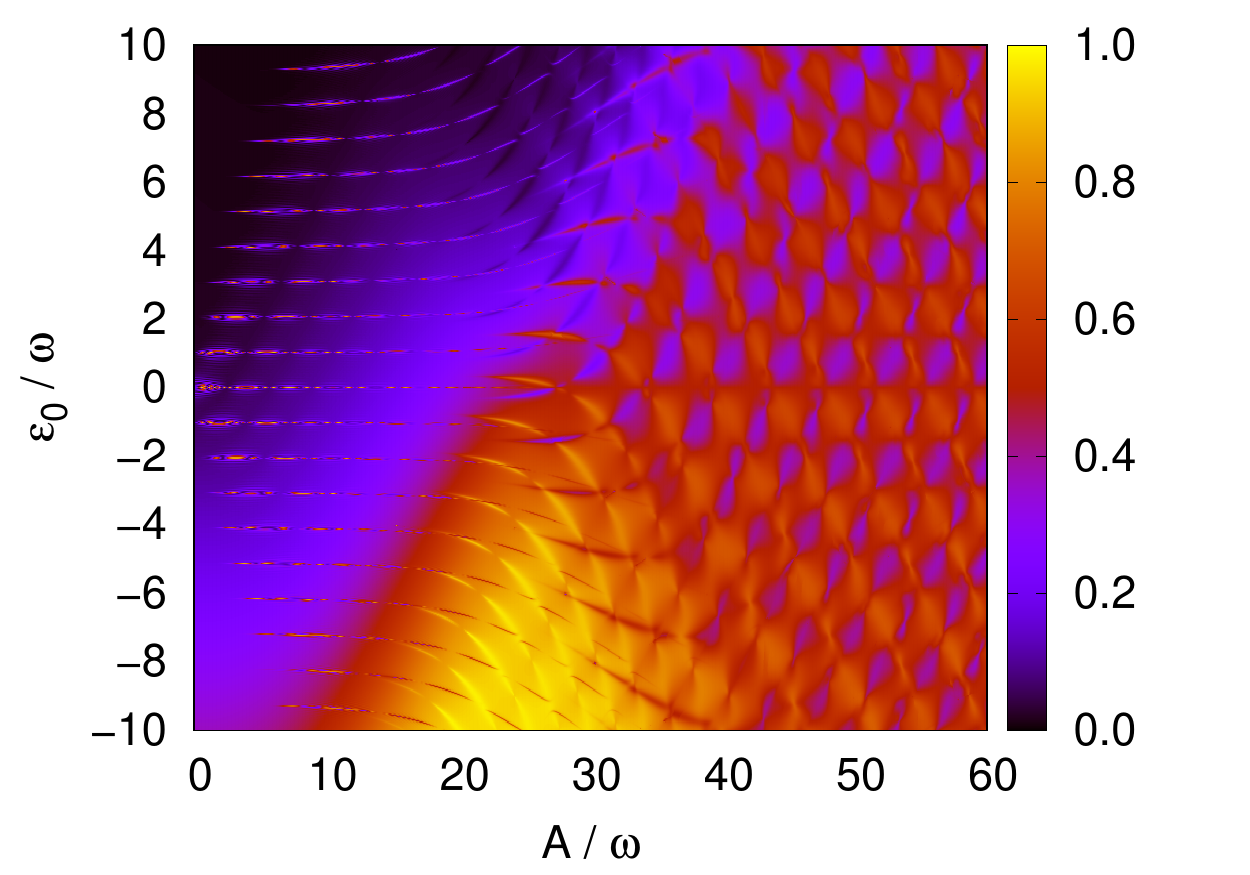}
			\put(0,62){(a)}
		\end{overpic}
	\end{subfigure}
	\begin{subfigure}[ht!]{0.5\textwidth} 
		\centering
		\begin{overpic}[width=0.85\textwidth]{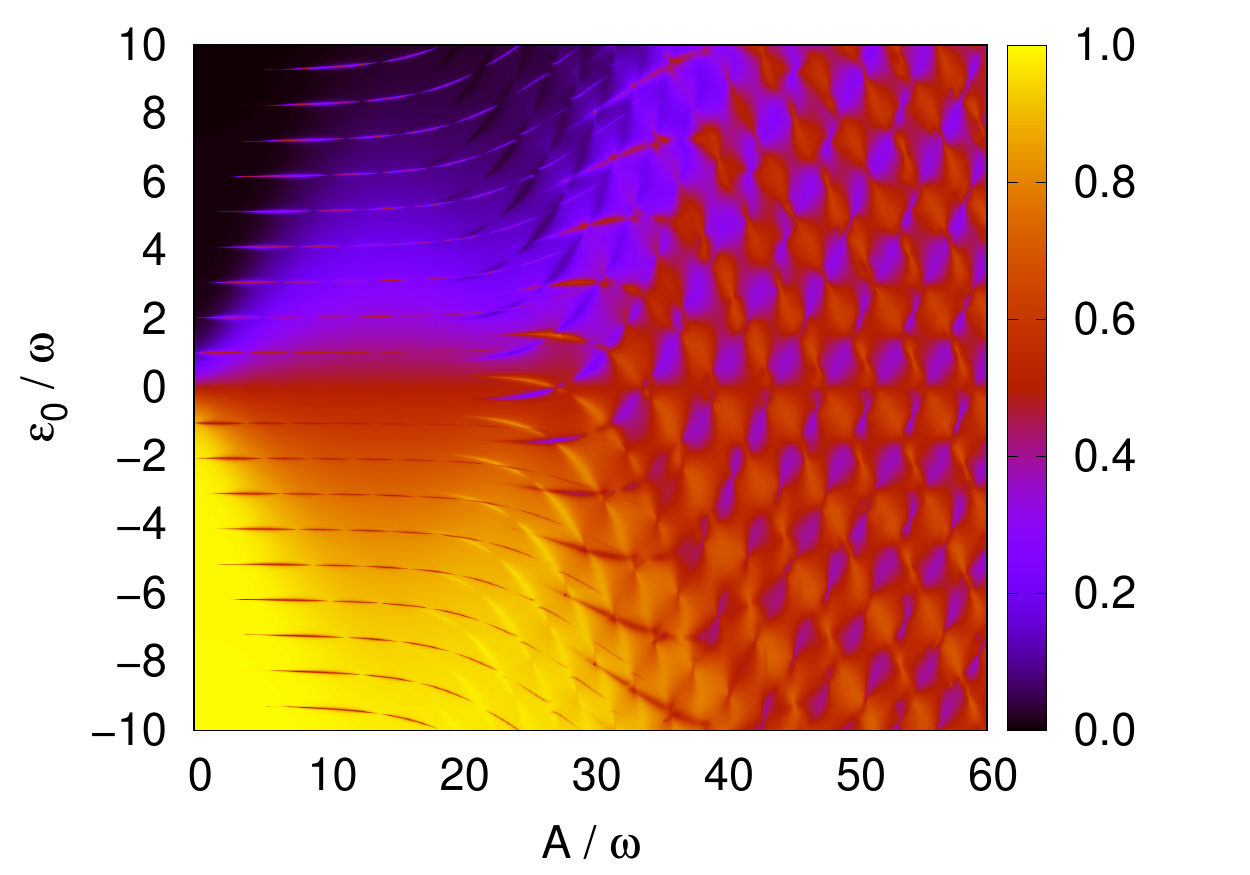}
			\put(0,62){(b)}
		\end{overpic}
	\end{subfigure}
	\caption{LZS interference patterns for the Rabi Hamiltonian in the USC Regime ($g/\omega_r=0.1125$)  considering effects of dissipation. Plots of $\overline{P}_{\ket{\uparrow}}(t)$ as a function of the driving parameters $A$ and $\varepsilon_0$, at finite time $t=50\tau$ (a) and in the asymptotic regime $t=\infty$ (b). The calculations were performed for $\omega/\omega_r=0.0375$, $\Delta/\omega_r=0.0038$, $T=0.0175\,\omega_r/k_B$ and $\kappa=0.001$ (see text for details).}
	\label{fig:8}
\end{figure}

One can also compare with the LZS pattern of Ref. \onlinecite{ferron_2016}, where a qubit  coupled to  a structured bath through $\sigma_z$ was analyzed.
We can identify  differences and similarities with  the present work: i) The first diamond LZS pattern of Ref. \onlinecite{ferron_2016} presents anti-symmetric resonances (characteristic of longitudinal coupling to the bath) instead of the narrow resonances and background observed here in Region III of Fig.\ref{fig:5} (identified with the first diamond) ii) The second diamond of the LZS pattern of Ref. \onlinecite{ferron_2016}  presents the same structure of alternating bright and dark lobes obtained here in Region VI, which has been explained in terms of red and blue sidebands.

The case of ultra strong coupling ($g=0.1125\,\omega_r$, shown in Fig.\ref{fig:8}) cannot be reduced to the ``qubit + structured bath" picture. In this situation the photonic gaps $\Delta_n=2 g\sqrt{n+1}$ are larger than the driving frequency $\omega$, and the associated photonic-LZS patterns are in the slow driving regime, as discussed in the previous section. 
In Fig.\ref{fig:8}(a) we plot the LZS pattern at a finite time $t=50\tau$.
At short times it is possible to observe the effect of the initial condition (the state $\ket{\downarrow, 0}$), and the plot resembles  the results obtained for the unitary evolution in Fig.\ref{fig:4}(b).
The steady state regime is shown in  Fig.\ref{fig:8}(b). The first diamond, in Region III, is similar to the one in Fig.\ref{fig:6}(b), since it corresponds to the qubit-LZS pattern.
The different behavior of the USC regime is manifested for amplitudes $A$ beyond the first diamond. In Region IV one can distinguish the arc-shaped photonic resonances with the arcs centered around the point $A=0$ , $\varepsilon_0=-\omega_r$. Similarly, in Region V one can observe the arcs centered around the point $A=0$, $\varepsilon_0=\omega_r$, corresponding to the other photonic resonances. More interestingly, in the second diamond, Region VI, the combined pattern of the inter crossing of  two arc-shaped photonic resonances is observed. This later  intercrossed pattern structure is an interested  and novel signature of the driven cQED in the USC regime.
As a final comment we stress that for the USC regime, the steady state  is attained for time scales shorter than in the SC regime. In our case for $t=1000\tau$ the LZS patterns in the USC (not shown) resemble those of the stationary case obtained in Fig.\ref{fig:8}(b).

\section{Concluding Remarks}
\label{sec:IV}

To summarize, we have  thoroughly analized  the LZS interference patterns that arise in a realistic cQED architecture taking into account the noise effects introduced by the environment, i.e. decoherence and relaxation. We studied the system composed by a harmonically driven superconducting qubit that is transversally coupled to a transmission line resonator.

We considered different values of the qubit-resonator coupling strength corresponding to the Strong Coupling (SC) and Ultra Strong Coupling (USC) regimes and observed important differences in the resonance patterns between both situations. A comprehensive description of the results was given in terms of the energy spectrum of the system Hamiltonian. We analyzed how the environment affects the  LZS patterns for different time scales and compared these results with those obtained when noise is neglected. 

We identify in the  LZS patterns the contributions   due to the qubit gap at $\varepsilon_0=0$ and those  due to the photonic gaps at  $\varepsilon_0=\pm\omega_r$. In particular, it was shown that for large amplitudes the interference patterns can be interpreted as the combined intercrossing of patterns of qubit -LZS and photonic-LZS.

Dissipative effects induce dramatic changes in the structure of the LZS patterns in comparison to the ideal (noiseless) case. The features  analyzed along this work
could help to design better strategies to mitigate noise  in LZS interferometry,
opening the possibility to extend the field of cQED for the case of strongly driven qubits. 

We acknowledge  financial support from CNEA, CONICET (PIP11220150100756), UNCuyo (P 06/C455) and ANPCyT (PICT2014-1382, PICT2016-0791).

\renewcommand\thefigure{\thesection.\arabic{figure}}

\bibliography{references}

\end{document}